\documentclass[apj,twocolumn]{emulateapj}
\usepackage{apjfonts}
\newcommand{\beq}{\begin{equation}}
\newcommand{\eeq}{\end{equation}}
\newcommand{\bea}{\begin{eqnarray}}
\newcommand{\eea}{\end{eqnarray}}

\begin{document}
\title{Shock Vorticity Generation from Accelerated Ion Streaming in the Precursor of Ultrarelativistic Gamma-Ray Burst External Shocks}
\author{Sean M. Couch\altaffilmark{1}, Milo\v s Milosavljevi\'c\altaffilmark{1}, and Ehud Nakar\altaffilmark{2}}
\altaffiltext{1}{The University of Texas, 1 University Station C1400, Austin, TX  78712}
\altaffiltext{2}{Theoretical Astrophysics, Mail Code 130-33, California Institute of Technology, 1200 East California Boulevard, Pasadena, CA 91125.  }
\righthead{ACCELERATED ION STREAMING IN GRB SHOCK PRECURSORS}
\lefthead{COUCH, MILOSAVLJEVI\'C, \& NAKAR}

\begin{abstract}

We investigate the interaction of nonthermal ions (protons and nuclei) accelerated in an ultrarelativistic blastwave with the pre-existing magnetic field of the medium into which the blastwave propagates.  While particle acceleration processes such as diffusive shock acceleration can accelerate ions and electrons, the accelerated electrons suffer larger radiative losses.  Under certain conditions, the ions can attain higher energies and reach farther ahead of the shock than the electrons, and so the nonthermal particles can be partially charge-separated.  To compensate for the charge separation, the upstream plasma develops a return current, which, as it flows across the magnetic field, drives transverse acceleration of the upstream plasma and a growth of density contrast in the shock upstream. If the density contrast is strong by the time the fluid is shocked, vorticity is generated at the shock transition. The resulting turbulence can amplify the post-shock magnetic field to the levels inferred from gamma-ray burst afterglow spectra and light curves. Therefore, since the upstream inhomogeneities are induced by the ions accelerated in the shock, they are generic even if the blastwave propagates into a medium of uniform density.  We speculate about the global structure of the shock precursor, and delineate several distinct physical regimes that are classified by an increasing distance from the shock and, correspondingly, a decreasing density of nonthermal particles that reach that distance.


\end{abstract}

\section{Introduction}
\label{sec:intro}

The standard model for gamma-ray burst (GRB) afterglow emission requires nonthermal acceleration of electrons to ultrarelativistic energies.  This can take place in the ``external'' relativistic shock wave driven into the circumburst medium by the burst ejecta \citep[e.g.,][and references therein]{Piran:05a,Meszaros:06}.  Candidate mechanisms for shock acceleration, such as the first order Fermi mechanism---also known as diffusive shock acceleration (DSA; e.g., \citealt{Bell:78,Blandford:78,Blandford:87})---can accelerate all charged particles, including the positively charged ions, for whose acceleration in GRB afterglows we still do not have direct observational evidence.  While theoretical models for particle acceleration in shocks have largely been based on analytic or semianalytic prescriptions for particle injection into the acceleration cycle and its scattering in the random magnetic field of the shock \citep[e.g.,][and references therein]{Achterberg:01,Ellison:02,Niemiec:06}, recent two-dimensional first-principles particle-in-cell simulations of unmagnetized pair shocks have delivered the first encouraging direct evidence for the formation of the signature nonthermal power-law tail \citep{Keshet:08,Spitkovsky:08}.  The power law tail seems to arise in tandem with a self-sustained electromagnetic structure in the shock precursor, where the nonthermal particles that move ahead of the shock excite plasma instabilities.  In their nonlinear development, the instabilities generate a magnetic field that scatters the nonthermal particles back toward the shock and thus closes the acceleration cycle.  Our aim here is to explore the consequences of the formation of a power-law tail of accelerated particles for shock hydrodynamics when a weak, large scale magnetic field is present in the shock upstream.

Inverse Compton cooling of the nonthermal electrons imposes an upper limit on the energy of electrons that are accelerated by DSA \citep[e.g.,][]{Li:06}.  Since higher energy ions reach farther from the shock into the shock upstream, there may exist a region in the shock upstream devoid of nonthermal electrons that still contains nonthermal ions; we restrict our analysis to this region.  In the reference frame in which the shock upstream is at rest, the nonthermal ions are beamed forward within the narrow angle $\sim \Gamma^{-1}$, where $\Gamma\gg 1$ is the shock Lorentz factor.  The upstream plasma responds to the forward-beamed nonthermal ions by developing a return current \citep[see, e.g.,][and references therein]{Thode:75,Spicer:84} which flows toward the shock.  The return current cancels the charge separation between the nonthermal ions that reach farther from the shock and the nonthermal electrons that are confined closer to the shock by the inverse Compton losses.  If the upstream plasma is weakly magnetized, the flow of return current across a weak magnetic field accelerates the upstream plasma transversally, i.e., perpendicular to the direction of shock propagation, which creates inhomogeneities in the plasma \citep{Reville:06,Pelletier:08}; this is a relativistic version of the mechanism discussed by \citet{Bell:04,Bell:05}.

We show that even when the upstream magnetic field is  weak, these density inhomogeneities may be sufficient to generate a stronger magnetic field in the shock downstream via vorticity production at the shock transition \citep{Goodman:08,Sironi:07}. This process converts the bulk kinetic energy of the shock downstream into vortical and magnetic energy. An advantage of this process is that it can take place even when the circumburst medium is uniform and therefore it is relevant for, as required by observations, both long and short GRBs \citep{Nakar:07}. Alternative sources of vorticity that were previously discussed in the literature are pre-existing density inhomogeneities in situations where an ultrarelativistic blastwave propagates into a clumpy stellar wind \citep{Sironi:07}, or an initially anisotropic blastwave \citep{Milosavljevic:08}.  Nonthermal particle streaming may generate or amplify a magnetic field in the shock upstream (\citealt{Bell:01,Milosavljevic:06}; A. Spitkovsky, priv. comm., but also see \citealt{Pelletier:08}).  The relative efficiencies of the direct, nonthermal particle-driven field amplification in the shock upstream, and the amplification in the shock transition via vorticity generation in the presence of streaming-generated density inhomogeneity will depend on how the shock energy is partitioned between the thermal fluid and the nonthermal accelerated particles and on the efficiency with which magnetic field is generated by cosmic ray-driven turbulence in the shock upstream.

Throughout this work we assume that in ultraretivistic collisionless shocks, DSA accelerates ions to high nonthermal energies.  While there is direct observational evidence of electron acceleration in such shocks, the assumption that protons are accelerated also may not necessarily be justified and deserves further investigation. Specifically, on theoretical grounds, particle acceleration should be inefficient when a compressed oblique large scale upstream magnetic field dominates particle dynamics in the shock downstream. In this case shock compression leads to a downstream magnetic field that is almost perpendicular to the shock normal. Then, a particle with a gyroradius smaller than the coherence length of the field cannot complete two full DSA cycles across the shock before becoming irreversibly bound to and escaping into the shock downstream \citep[e.g.,][]{Lemoine:06b}. Therefore, a field that contains large-amplitude magnetic field perturbations on small scales seems to be required for acceleration \citep[e.g.,][]{Ostrowski:02,Niemiec:04,Niemiec:06,Niemiec:06b}. Even in the conditions in which a shock propagates in a weakly magnetized medium and generates magnetic turbulence that serves as a scattering agent for energetic particles, it cannot be taken for granted that the resulting particle spectrum will extend up to the high Lorentz factors that are required for the mechanism that we present. These potential limitations should be born in mind in what follows.

This work is organized as follows.  In \S~\ref{sec:charge_separation} we analyze charge separation in the shock precursor and relate the nonthermal ion density ahead of the shock to the maximum energy to which nonthermal ions are accelerated in the shock.  In \S~\ref{sec:eom} we study the motion of the upstream medium in response to the Lorentz force associated with the nonthermal ion streaming across a weak magnetic field.  In \S~\ref{sec:dencon} we determine the conditions under which nonlinear density inhomogeneities can result and estimate the energy density of the magnetic field produced by vortical motions that are excited as the shock transition sweeps past the inhomogeneities.  In \S~\ref{sec:bell} we examine the production of a small scale reversing magnetic field structure that was assumed in \S~\ref{sec:eom}, in the weak large-scale magnetic field of the circumburst medium by the nonresonant streaming instability introduced by \citet{Bell:04,Bell:05}. In \S~\ref{sec:discussion} we speculate about the global structure of the shock precursor, and delineate several distinct physical regimes that are classified by an increasing distance from the shock.  In \S~\ref{sec:conclusions}, we summarize our main conclusions.

\section{Interaction of the Shock Precursor with the Upstream Plasma}

\subsection{Charge Separation}
\label{sec:charge_separation}

We consider a relativistic blast wave described by Lorentz factor $\Gamma$ in which electrons and positive ions are accelerated by DSA. Unless otherwise explicitly noted, all quantities are defined in the rest frame of the upstream. The maximum energy to which the electrons are accelerated, $\gamma_{e,{\rm max}} m_e c^2$, is limited by inverse Compton cooling as the electrons traverse the shock upstream during a DSA cycle \citep[e.g.,][]{Li:06}. The ions experience much smaller radiative losses and can be accelerated to higher energies, $\gamma_{i,{\rm max}} m_i c^2$, than the electrons.  Here, $m_i$ is the ion mass, and $\gamma_e$ and $\gamma_i$ denote the electron and ion Lorentz factors, respectively.    Hence, the accelerated ions can travel farther ahead of the shock than the electrons. If $\gamma_{i,{\rm max}}m_i>\gamma_{e,{\rm max}}m_e$, the shock precursor consisting of accelerated particles will not be charge neutral far from the shock, at distances that the nonthermal electrons cannot reach.  There, charge neutrality is recovered by a small polarization of the upstream, unshocked plasma. The polarization is established by the flow of a return current in the upstream plasma  \citep[see, e.g.,][and references therein]{Thode:75,Spicer:84}.  This return current drives the growth of upstream plasma inhomogeneity \citep{Bell:04,Bell:05} that is the subject of this work.

The maximum distance $\Delta(\gamma)$ that a nontermal particle with Lorentz factor $\gamma$ can reach from the shock is determined by the magnetic field structure in the shock upstream.  The magnetic field can come from the weak pre-existing magnetization of the circumburst medium, or can be generated by plasma self-organization resulting from an interaction with the streaming of nonthermal particles.  Depending on the structure of the magnetic field, nonthermal particle orbits are either deflected in a circular fashion if the magnetic field is coherent, or execute random walks if the field is tangled, i.e., reverses on scales $\lambda$ smaller than the gyroradius of the nonthermal particles divided by the shock Lorentz factor, $r_{\rm g}/\Gamma\sim \gamma m c^2/e\bar B\Gamma$, where $\bar B$ is the RMS field strength.  Since the magnetic field can have different strength and structure on different scales $\bar B(\lambda)$, we employ the subscript ``def'' to denote the magnetic field length scale that produces the largest (circular or random) deflection of the nonthermal particle.

When a nonthermal particle is deflected by angle $\gtrsim \Gamma^{-1}$ from the shock normal, its velocity parallel to the direction of shock propagation becomes smaller than that of the shock and the shock can catch up with the particle \citep[see, e.g.,][and references therein]{Achterberg:01}.  A particle that has been overtaken by the shock can be scattered back into the shock upstream, as is necessary to close the DSA cycle.  Therefore, working in the reference frame in which the shock upstream is at rest, for the coherent field we have $\Delta(\gamma)\sim r_{\rm g}/\Gamma^3$, whereas for the tangled field we obtain $\Delta(\gamma)\sim r_{\rm g}^2/\lambda_{\rm def}\Gamma^4$, where $r_{\rm g}=\gamma m c^2/e\bar B_{\rm def}$.  Since we do not estimate the magnetic field strength and geometry from first principles, we parametrize the effect of the deflecting, possibly small scale magnetic field by the maximum Lorentz factor  of a nonthermal ion $\gamma_{p,{\rm max}}$, which we here take to be a proton, that can be accelerated with the aid of the magnetic field.  We separately treat the cases in which the deflecting field is coherent and tangled.  In the case of a tangled field, we assume that the field is tangled on spatial scales $\lambda_{\rm def}\ll r_{\rm g}/\Gamma$ that are independent of the distance from the shock $\Delta$. This assumption may not be valid in general \citep[see, e.g.,][]{Milosavljevic:06,Katz:07}, in which case our analysis should be interpreted as applying to the highest energy accelerated particles, and to the magnetic field component, with its own characteristic correlation length $\lambda_{\rm def}$,  that has the greatest influence on the dynamics of these particles.

We also assume that the blastwave propagates into a quasi-homogeneous medium with average density $\bar \rho\sim m_{\rm p} n$.  The highest energy protons can reach the farthest from the shock to a distance $\Delta_p(\gamma_{p,{\rm max}})\sim R/8\Gamma^2$, where $R$ is the radius of the blastwave, and the factor of one eighth is peculiar to adiabatic spherical ultrarelativistic blastwaves propagating into uniform density media \citep{Blandford:76}.  Let $U_{\rm sh}$ denote the energy density in radiation at radius $R$ in the shock frame which is a fraction $\epsilon_{\rm rad}=(4\pi/3) R^3 U_{\rm sh}/E_{\rm tot}$ of the total isotropic equivalent energy of the blastwave $E_{\rm tot}$.  Equating the inverse Compton power $\frac{4}{3} \sigma_{\rm T} c \gamma_{e,{\rm sh}}^2 U_{\rm sh}$, where $\gamma_{e,{\rm sh}}\sim \gamma_e/\Gamma$ is the electron Lorentz factor in the shock frame, $\sigma_{\rm T}$ is the Thompson cross section, and we neglected the Klein-Nishina effects \citep[e.g.,][]{Li:06}, to the energy gain $\sim \gamma_{e,{\rm sh}} m_e c^2$ per DSA cycle of duration $\sim \Delta_{e,{\rm sh}}/c \sim \Gamma\Delta_e/c$, we obtain the maximum  Lorentz factor to which electrons can be accelerated, in spite of the cooling, as a function of the Lorentz factor to which the non-cooling protons can be accelerated
\bea
\label{eq:gamma_e_max}
\gamma_{e,{\rm max}}
 &\sim&
\frac{4.4\ c^{1/3} m_p^{1/2} \gamma_{p,{\rm max}}^{1/2}\Gamma^{1/3}}{\epsilon_{\rm rad}^{1/2}\bar\rho^{1/3} E_{\rm tot}^{1/6} \sigma_{\rm T}^{1/2} } \textrm{ (coherent field) } , \nonumber\\
&\sim&
\frac{2.7\ c^{2/9} m_p^{2/3} \gamma_{p,{\rm max}}^{2/3}\Gamma^{2/9}}{\epsilon_{\rm rad}^{1/3}\bar\rho^{2/9} E_{\rm tot}^{1/9} m_e^{1/3} \sigma_{\rm T}^{1/3} } \textrm{ (tangled field) } ,
\eea
which, as usual, is expressed in the rest frame of the shock upstream.  Here, we have eliminated the blastwave radius $R$ via the relation $R=(17 E_{\rm tot}/8\pi\Gamma^2 \bar\rho c^2)^{1/3}$ applicable to an adiabatic spherical blastwaves propagating into uniform density media \citep{Blandford:76}.  The magnetic field length scale $\lambda_{\rm def}$ (in the case of a tangled field) and strength squared $B_{\rm def}^2$ do not appear in equation (\ref{eq:gamma_e_max}) because they have been expressed in terms of the maximum proton Lorentz factor that can be accelerated by deflection in a field with these properties via $B_{\rm def} =[(\gamma_{p,{\rm max}} m_p c^2 /e)/\Gamma^3]/(R/8\Gamma^2)$ (coherent field) and $\lambda_{\rm def} B_{\rm def}^2 =[(\gamma_{p,{\rm max}} m_p c^2 /e)^2/\Gamma^4]/(R/8\Gamma^2)$ (tangled field).

The nondimensional distance from the shock that the most energetic electrons can reach, relative to the distance that the protons can reach, can be expressed as
\bea
\label{eq:Delta_ratio}
x_{\rm cool}&\equiv&
\frac{\Delta_e(\gamma_{e,{\rm max}})}{\Delta_p(\gamma_{p,{\rm max}})} \nonumber\\&\sim&
\frac{4.4\ c^{1/3} m_e \Gamma^{1/3}}{ \gamma_{p,{\rm max}}^{1/2}\epsilon_{\rm rad}^{1/2}\bar\rho^{1/3} E_{\rm tot}^{1/6} m_p^{1/2} \sigma_{\rm T}^{1/2} } \textrm{ (coherent field), } \nonumber\\ &\sim&\
\frac{7\ c^{4/9} m_e^{4/3} \Gamma^{4/9}}{ \gamma_{p,{\rm max}}^{2/3}\epsilon_{\rm rad}^{2/3}\bar\rho^{4/9} E_{\rm tot}^{2/9} m_p^{2/3} \sigma_{\rm T}^{2/3} } \textrm{ (tangled field) } .
\eea
For typical values of $\epsilon_{\rm rad}$, $E_{\rm tot}$, $\Gamma$, and $n$, this becomes
\bea
\label{eq:x_cool}
x_{\rm cool}&\sim&
\left(\frac{\gamma_{p,{\rm max}}}{460}\right)^{-1/2}
\left(\frac{\epsilon_{\rm rad}}{0.1}\right)^{-1/2}
\left(\frac{E_{\rm tot}}{10^{53}\textrm{ erg}}\right)^{-1/6}
\nonumber\\ & &\times
\left(\frac{\Gamma}{100}\right)^{1/3}
\left(\frac{n}{1\textrm{ cm}^{-3}}\right)^{-1/3}  \textrm{ (coherent field), }\nonumber\\
&\sim&
\left(\frac{\gamma_{p,{\rm max}}}{440}\right)^{-2/3}
\left(\frac{\epsilon_{\rm rad}}{0.1}\right)^{-2/3}
\left(\frac{E_{\rm tot}}{10^{53}\textrm{ erg}}\right)^{-2/9}
\nonumber\\ & &\times
\left(\frac{\Gamma}{100}\right)^{4/9}
\left(\frac{n}{1\textrm{ cm}^{-3}}\right)^{-4/9}  \textrm{ (tangled field), }
\eea
which shows that for $\gamma_{p,{\rm max}}\gg 10^3$, regardless of the magnetic field geometry, the electrons are unable to travel as far from the shock as the protons and $x_{\rm cool}\ll 1$ in either case.  Note that for a coherent confining field
\bea
\gamma_{p,{\rm max}} &\sim& 7 \times10^5
\left(\frac{\bar B_{\rm def}}{10^{-6}}\right)
\left(\frac{E_{\rm tot}}{10^{53}\textrm{ erg}}\right)^{1/3}
\nonumber\\&& \times
\left(\frac{\Gamma}{100}\right)^{1/3}
\left(\frac{n}{1\textrm{ cm}^{-3}}\right)^{1/3} \textrm{ (coherent field)},
\eea
implying that if the confinement is dominated by a pre-existing microgauss upstream magnetic field we expect that $\gamma_{p,{\rm max}} \sim 10^5-10^6$, while a larger value is expected if the shock precursor generates a magnetic field in which the tangled component dominates particle deflection. Therefore we always expect that $x_{\rm cool}\ll 1$. In what follows, we ignore the interior region $\Delta < \Delta_e(\gamma_{e,{\rm max}})$ populated by nonthermal particles of either sign and assume that the exterior region $\Delta > \Delta_e(\gamma_{e,{\rm max}})$ contains only nonthermal ions, which we have assumed to be protons.

The nonthermal particle concentration at a given distance ahead of the shock will be dominated by the lowest energy particles that reach that distance. Thus, we can relate the nonthermal proton density $n_{\rm ntp}$ in the shock upstream to their energy distribution
\beq
\label{eq:crdens}
n_{\rm ntp} (\gamma) \sim \frac{1}{4 \pi R^2 \Delta (\gamma)} \frac{dN^{({\rm up})}_{\rm ntp}}{d \ln \gamma} ,
\eeq
where $N^{({\rm up})}_{\rm ntp}(\gamma)$ is the total number of nonthermal protons in the shock upstream with Lorentz factors less than $\gamma$.

We assume that the accelerated particle spectrum in the shock downstream resulting from DSA contains equal energy on all energy scales, $dN_{\rm ntp}^{({\rm down})}/d\gamma \propto \gamma^{-2}$ \citep[cf. e.g.,]{Lemoine:03,Ellison:02,Ellison:04}; this assumption is not critical as somewhat steeper spectra, expected when particle scattering in the shock downstream is not isotropic \citep[e.g.,][]{Keshet:05,Lemoine:06a}, lead to similar conclusions.  Ignoring the energy in nonthermal protons located in the upstream, we may normalize the  spectrum by requiring that
\beq
\label{eq:enint}
\int_{\gamma_{p,{\rm min}}}^{\gamma_{p,{\rm max}}} \gamma m_p c^2 \frac{dN_{\rm ntp}^{({\rm down})}}{d\gamma} d\gamma = \epsilon_{\rm nt} E_{\rm tot} ,
\eeq
where $\epsilon_{\rm nt}$ is the fraction of the total energy in the accelerated protons.  The minimum Lorentz factor of the nonthermal particles, resulting from a single scattering across the shock is $\gamma_{p,{\rm min}} \sim \Gamma^2$ \citep{Gallant:99}, but the dependence on $\gamma_{p,\rm min}$ is only logarithmic.

To relate $N^{({\rm up})}_{\rm ntp}$ to $N^{({\rm down})}_{\rm ntp}$, let $P\sim \frac{1}{2}$ denote the probability that a particle in the shock upstream, upon returning to the shock, is scattered again into the shock upstream.  The downstream then contains the particles that are not re-scattered into the upstream, but are entrapped within and are advecting with the downstream.  Conservation of nonthermal proton flux at the shock transition then implies\footnote{This corrects the inadequate assumption $N^{({\rm up})}_{\rm ntp}(\gamma) \sim N^{({\rm down})}_{\rm ntp}(\gamma)$ that we had made in \citet{Milosavljevic:06}.}
\beq
\frac{dN^{({\rm down})}_{\rm ntp}}{d\ln\gamma} \sim (1-P) \frac{t_{\rm dyn}}{t_{\rm acc}(\gamma)} \frac{dN^{({\rm up})}_{\rm ntp}}{d\ln\gamma} .
\eeq
Here $t_{\rm acc}$ is the time between the emission of a nonthermal proton from the shock and its reabsorption in the shock
\beq
t_{\rm acc}(\gamma) \sim \frac{(1-\bar\beta_{\parallel,{\rm sh}}^2)^{1/2}}{\bar\beta_{\parallel,{\rm sh}}} \frac{\Gamma^2\Delta(\gamma)}{c} ,
\eeq
where $\bar\beta_{\parallel,{\rm sh}}\sim \frac{1}{2}$ is the average component parallel to the direction of shock propagation of the velocity of nonthermal protons in the shock frame, and $t_{\rm dyn}\sim R/c$ is the dynamical age of the blastwave. From this,
\bea
\label{eq:N_up}
\frac{dN^{({\rm up})}_{\rm ntp}}{d\ln\gamma} \sim \frac{1}{\sqrt{8}} \left(\frac{\gamma}{\gamma_{p,{\rm max}}}\right)^s \frac{dN^{({\rm down})}_{\rm ntp}}{d\ln\gamma} ,
\eea
where $s\equiv d\ln\Delta/d\ln\gamma$, so that $s=1$ for the coherent field, and $s=2$ for the tangled field.  This shows that the integral spectrum of all the nonthermal particles located in the shock upstream is substantially harder than that in the downstream.

Substituting equation (\ref{eq:N_up}) in equation (\ref{eq:crdens}), we find that the total density of the accelerated protons equals
\bea
\label{eq:density_of_delta}
n_{\rm ntp}(\Delta) &\sim&
 \frac{0.04 \epsilon_{\rm nt} \bar\rho^{2/3}E_{\rm tot}^{1/3}\Gamma^{4/3}}{\Delta m_p c^{2/3} \gamma_{p,{\rm max}}  \Lambda} \textrm{ (coherent field) },
\nonumber\\ &\sim&  \frac{0.1\ \epsilon_{\rm nt} \bar\rho^{5/6}E_{\rm tot}^{1/6}\Gamma^{8/3}}{\Delta^{1/2}m_p c^{1/3} \gamma_{p,{\rm max}} \Lambda} \textrm{ (tangled field) },
\eea
where again $\Delta(\gamma)=(\gamma/\gamma_{p,{\rm max}})^s R/8\Gamma^2$ and we have defined $\Lambda\equiv\ln(\gamma_{p,{\rm max}}/\gamma_{p,{\rm min}})\sim 5-20$.  Let $x\equiv \Delta / (R/8\Gamma^2)$ denote the nondimensionalized distance from the shock; in terms of this parameter,
\bea
\label{eq:ntp_density}
n_{\rm ntp}(x)
&\sim&
\frac{0.3}{x^q} \left(\frac{\gamma_{p,{\rm max}}}{10^6}\right)^{-1}\left(\frac{\Gamma}{100}\right)^4
\left(\frac{\bar n}{1\textrm{ cm}^{-3}}\right) \left(\frac{\epsilon_{\rm nt}}{0.1}\right) ,
\eea
where $q=s^{-1}$, so that $q=1$ for the coherent field and $q=\frac{1}{2}$ for the tangled field, and we have set $\Lambda=10$.  Equation (\ref{eq:ntp_density}) gives the nonthermal proton density in the rest frame of the shock upstream.  For large $\Gamma$ or small $\gamma_{p,{\rm max}}$, this density can formally exceed the upstream density; in reality, $n_{\rm ntp}$ may be limited by kinetic plasma instabilities.  In the rest frame of the nonthermal proton shock precursor, the density ratio $n_{\rm ntp}/\bar n$ is reduced by factor $\Gamma^{-2}$.

\subsection{Dynamics}
\label{sec:eom}

Here we study the response of the shock upstream to the streaming of the nonthermal protons in the shock precursor.  The streaming of nonthermal protons with density $n_{\rm ntp}$ will give rise to a return current
\beq
{\bf J}_{\rm ret}(\Delta)\sim  - ecn_{\rm ntp}(\Delta) \hat{\bf R} ,
\eeq
where $\hat{\bf R}$ is a unit vector in the direction of propagation of the shock.  The return current cancels the charge separation among the nonthermal particles resulting from the inverse Compton cooling suppression of the acceleration of nonthermal electrons. Consider a fluid element in the upstream medium containing a magnetic field ${\bf B}$ that reverses direction on length scales $\lambda$. This may not be the same magnetic field scale that is the most effective at deflecting the orbits of the nonthermal particles, see \S~\ref{sec:charge_separation}. A magnetic field that reverses on the relevant scales may exist in the upstream at the outset.  If not, as we show in \S~\ref{sec:bell}, it can be produced by the return current from an initially coherent microgauss field. Let $\mu$ denote the sine of the angle between the magnetic field and the direction of shock propagation.

As the return current flows across the field, the plasma experiences a Lorentz force acting perpendicular to both the direction of shock propagation and the field direction.  Let $\tau$ denote the time measured since the arrival of the photon shell; in this time coordinate, the shock transition sweeps a fluid element at $\tau = R/8\Gamma^2c$. The equation of motion for the transverse Lagrangian displacement of the fluid element in response to the Lorentz force, valid while the motion is Newtonian, i.e., while $|d\vec\xi/d\tau|\ll c$, reads
\bea
\label{eq:eom}
\frac{d^2 {\vec \xi}}{d \tau^2} &=& \frac{{\bf J}_{\rm ret}(\tau) \times {\bf B} }{\rho c} .
\eea
From this, assuming that ${\bf B}\sim B \sin(2\pi {\bf r}\cdot\hat{\bf B}/\lambda)\hat{\bf B}$, where ${\bf r}$ is a spatial coordinate and $B$ is peak field, the amplitude of the transverse acceleration is
\beq
\frac{d^2\xi }{d\tau^2}\sim
\frac{1.7\ \mu e c \epsilon_{\rm nt} \epsilon_B^{1/2}\bar\rho^{1/2} \Gamma^4}{m_p \gamma_{p,{\rm max}} \Lambda } x^{-q}\sin\phi,
\eeq
where $\phi\equiv 2\pi {\bf r}\cdot\hat{\bf B}/\lambda$, and $\epsilon_B\equiv B^2/8\pi\bar\rho c^2$ is the energy density in the magnetic field on scales $\lambda$ divided by the rest energy density of the shock upstream. Note that $\epsilon_B$ depends on the length scale $\lambda$ and is not directly related to the confining magnetic field energy density measured on scales $\lambda_{\rm def}$.

We define $x\equiv \Delta / (R/8\Gamma^2)$ so that the unperturbed motion of an upstream fluid element toward the shock follows the characteristic $dx/d\tau=-c/(R/8\Gamma^2)$. Ignoring the amplification of the magnetic field during the linear phase of the nonthermal proton-driven streaming instability (\S~\ref{sec:bell}) we obtain
\beq
\label{neweom}
\frac{d^2 \xi}{dx^2} \sim \frac{ 0.02\ \mu e \epsilon_B^{1/2} E_{\rm tot}^{2/3} \epsilon_{\rm nt} }{ \Gamma^{4/3} \bar n^{1/6} m_p^{7/6} c^{7/3} \gamma_{p,{\rm max}} \Lambda } x^{-q} \sin\phi.
\eeq
Given that the upstream fluid starts from rest, $\xi(1) = d\xi/dx(1) =0$, equation (\ref{neweom}) can be integrated to find the transverse displacement
\bea
\label{eq:fluid_y}
\xi(x) &\sim& \frac{0.02\ \mu e \epsilon_B^{1/2} E_{\rm tot}^{2/3} \epsilon_{\rm nt} }{ \Gamma^{4/3} \bar n^{1/6} m_p^{7/6} c^{7/3} \gamma_{p,{\rm max}} \Lambda} g(x)  \sin\phi .
\eea
where $g(x)=x\ln x-x+1$ for the coherent field and $g(x)=\frac{4}{3}x^{3/2}-2x+\frac{2}{3}$ for the tangled field.
Setting  $\mu=1$ and $\Lambda=10$ we obtain
\bea
\label{eq:xi_numerical}
\xi &\sim&
\frac{R}{8\Gamma}
\left(\frac{\gamma_{p,{\rm max}}}{1.6\times10^4}\right)^{-1}
\left(\frac{\epsilon_B}{10^{-9}}\right)^{1/2}
\left(\frac{E_{\rm tot}}{10^{53}\textrm{ erg}}\right)^{1/3}
\left(\frac{\epsilon_{\rm nt}}{0.1}\right)
\nonumber\\ & &\times
\left(\frac{\Gamma}{100}\right)^{1/3}
\left(\frac{\bar n}{1\textrm{ cm}^{-3}}\right)^{1/6} g(x) \sin\phi .
\eea
Thus for $\gamma_{p,{\rm max}}\gg 10^4$, the distance traveled by the upstream fluid is $\ll R/8\Gamma$, which is the width of the blastwave in the frame of the shock downstream. The value of $\xi(x=0)$ decreases with increasing $\gamma_{p,{\rm max}}$ because when protons are accelerated to higher energies, the number of nonthermal protons between which the available energy $\sim \epsilon_{\rm nt}E_{\rm tot}$ is partitioned decreases, while the lower energy nonthermal protons are confined closer to the shock, both of which decrease the density of nonthermal protons far from the shock where the Amp\`ere force-driven acceleration takes place.

Under what circumstances will the transverse displacement result in a nonlinear density contrast on scales $\lambda$?  This requires $\xi(x) \sim \lambda$ before the fluid element is overtaken by the shock at $x=0$.  However, before the fluid arrives at the shock transition, some assumptions entering the derivation of equation (\ref{eq:fluid_y}) may be compromised: its motion may become relativistic ($d\xi/d\tau\sim c$), it may enter the region in which the pressure transfered by the nonthermal protons to the upstream fluid starts to accelerate the upstream fluid in the original rest frame of the shock upstream, and it may also enter the region containing accelerated nonthermal electrons where the return current vanishes.  The latter problem arises when $x \sim x_{\rm cool}$ where $x_{\rm cool} \equiv \Delta_e(\gamma_{e,{\rm max}}) / (R/8\Gamma^2)$ is given in equation (\ref{eq:Delta_ratio}). It is evident from equation (\ref{eq:x_cool}) that the return current vanishes only in the region populated by the lowest-energy nonthermal protons ($\gamma_p\lesssim 10^3$), which for $\gamma_{p,{\rm max}}\gg 10^3$, as expected, will be confined very close to the shock transition. We are not able determine the value of $x$ at which the upstream starts to be accelerated in its own frame but we do assume that it is smaller than $x_{\rm cool}$.

To assess the distance at which the transverse motion of the fluid becomes relativistic, we focus on the special case of a tangled deflecting field; the case of a coherent deflecting field is qualitatively similar.
The time at which the upstream motion becomes relativistic is obtained by setting $d\xi/dx = -R/8\Gamma^2$ (i.e., $d\xi/d\tau = c$) to find
\bea
\label{eq:xirel}
x_{\rm rel}&=&\left(1-\frac{\gamma_{p,{\rm max}}}{\gamma_{\rm rel,crit}} \right)^{2} \textrm{ (tangled field) } ,
\eea
for $\gamma_{p,{\rm max}}\leq \gamma_{\rm rel,crit}$, where
\bea
\label{eq:gamma_rel_crit}
\gamma_{\rm rel,crit}&\sim&
\frac{0.4\ \Gamma^{4/3} \bar n^{1/6} \mu e \epsilon_B^{1/2} E_{\rm tot}^{1/3} \epsilon_{\rm nt} } {c^{5/3} m_p^{5/6}  \Lambda} \sin\phi\nonumber\\
&\sim&
3\ \times10^6
\left(\frac{\epsilon_B}{10^{-9}}\right)^{1/2}
\left(\frac{E_{\rm tot}}{10^{53}\textrm{ erg}}\right)^{1/3}
\left(\frac{\epsilon_{\rm nt}}{0.1}\right)
\nonumber\\ & &\times
\left(\frac{\Gamma}{100}\right)^{4/3}
\left(\frac{n}{1\textrm{ cm}^{3}}\right)^{1/6} \textrm{ (tangled field) } .
\eea
When $\gamma_{p,{\rm max}}>\gamma_{\rm rel,crit}$, the fluid does not attain relativistic transverse velocity before it reaches the shock. Then, the numerical estimate in equation (\ref{eq:xi_numerical}) at $x=0$ where $g(x) \sim 1$, which was derived assuming Newtonian motion, shows that the transverse displacement decreases with $\gamma_{p,{\rm max}}$.

Conversely, for $\gamma_{p,{\rm max}}\ll\gamma_{\rm rel,crit}$, the fluid motion formally becomes relativistic well before the fluid reaches the shock.   The maximum transverse displacement that can be attained in the Newtonian regime can be evaluated by substituting $x=x_{\rm rel}$ in equation (\ref{eq:fluid_y}) to obtain
\bea
\xi(x_{\rm rel}) &\sim& 0.3\ \frac{\gamma_{p,{\rm max}}m_p^{1/2} c \Lambda}{\mu e \epsilon_{\rm nt}\epsilon_B^{1/2} n^{1/2} \Gamma^4\sin\phi}\nonumber\\ &\sim&
0.004\
\frac{R}{8\Gamma} \left(\frac{\gamma_{p,{\rm max}}}{10^6}\right)
\left(\frac{\epsilon_B}{10^{-9}}\right)^{-1/2}
\nonumber\\ & &\times
\left(\frac{\epsilon_{\rm nt}}{0.1}\right)^{-1}
\left(\frac{E_{\rm tot}}{10^{53}\textrm{ erg}}\right)^{-1/3}
\left(\frac{\Gamma}{100}\right)^{-7/3}
\nonumber\\ & &\times
\left(\frac{n}{1\textrm{ cm}^{3}}\right)^{-1/6} \textrm{ (tangled field) } ,
\eea
for $\gamma_{p,{\rm max}}\ll \gamma_{\rm rel,crit}$. Acceleration of the fluid may not stop in the Newtonian regime; indeed, it will likely continue into the transrelativistic regime, where the equation of motion is simply $d\xi/d\tau\sim c$. Then, however, the motion will not be purely perpendicular to the direction of shock propagation and the fluid will start accelerate, in the upstream frame, in the parallel direction. Here we do not take into account this behavior and conservatively restrict our attention to the maximum transverse displacement that the fluid reaches while its motion remains Newtonian.

Since the transverse displacement increases with $\gamma_{p,{\rm max}}$ when the transverse motion becomes transrelativistic and decreases with $\gamma_{p,{\rm max}}$ when it does not, the largest Newtonian displacement is possible when the maximum proton Lorentz factor is about equal to $\gamma_{\rm rel,crit}$ in equation (\ref{eq:gamma_rel_crit}) and is simply given by $\xi_{\rm max}\sim \frac{1}{3}R/8\Gamma^2$, which is much larger than the proton plasma skin depth $c/\omega_{\rm p}\equiv c/(4\pi e^2 n/m_p)^{1/2}$ of the shock upstream.  Of course, the maximum transverse displacement is achieved only if a magnetic field that reverses on scales $\lambda\sim \xi_{\rm max}$ is present.  We speculate about the origin of the reversing field in \S~\ref{sec:bell} below. Note that $\xi_{max}$ is close to the maximal value allowed by causality, $R/8\Gamma^2$.

\subsection{Density Inhomogeneity, Vorticity, and Dynamo}
\label{sec:dencon}

Here, we study the evolution of an initially uniform density fluid in response to the Lorentz force-driven transverse motion discussed in \S \ref{sec:eom}.  If the magnetic field reverses itself on scales $\sim \lambda$, the average linear density contrast arising from the transverse motion, for linear displacements, $\xi\ll \lambda$, will be given by
\beq
\delta\equiv \left\langle\frac{\rho-\bar\rho}{\bar \rho}\right\rangle \sim  \langle \vec{\nabla}\cdot\vec{\xi}\rangle
\sim \frac{\langle\xi\rangle}{\lambda} .
\eeq
We expect nonlinear density contrast on scales $\lambda$ provided that $\xi(x_{\rm max}) \gtrsim \lambda$, where $x_{\rm max}={\rm max} \{x_{\rm cool},x_{\rm rel}\}$.  This condition places an upper limit on the magnetic field reversal length scales $\lambda$ on which a nonlinear density contrast can build up before the fluid reaches the shock.

When fluid with nonlinear density contrast reaches the shock transition, provided that the width of the density jump of the shock transition is smaller than the length scale associated with the contrast, vorticity is generated at the transition.  \citet{Sironi:07} have calculated  the fraction of shock energy that is converted into vortical energy for an ultrarelativistic blastwave propagating into a clumpy medium. They provide a fitting formula for vortical energy fraction $\epsilon_{\rm vort}$ as a function of clump size $L$, peak density contrast $\delta_{\rm max}$, and volume-filling factor $n_{\rm clump}L^3$, which reads
\beq
\epsilon_{\rm vort} \approx  \left(\frac{0.36\ n_{\rm clump} L^3}{\Gamma} \right ) \frac{ \delta_{\rm max}^2 }{1+0.176\ (\delta_{\rm max})^{1.054}} .
\eeq
For $L\sim\lambda$ and  $n_{\rm clump}\sim \lambda^{-3}$ and mildly nonlinear density contrast $\delta_{\rm max}\sim 1$, this implies $\epsilon_{\rm vort}\sim (3\Gamma)^{-1}$. The collisions of and compression in accelerated plasma shells \citep{Bell:04,Bell:05} may amplify the density contrast into the strongly nonlinear regime $\delta_{\rm max}\gg 1$, but we do not attempt to estimate the magnitude of this effect.

If vorticity is generated on scales $\lambda\ll R/\Gamma$, the resulting eddies can amplify the downstream magnetic field to levels $\epsilon_{B,{\rm down}}\sim \epsilon_{\rm vort}$ via a turbulent dynamo mechanism \citep{Meneguzzi:81,Schekochihin:04}.   Strong amplification is expected when the number of vortical eddy turnovers in the shock downstream is larger than unity, but this is automatically expected when the density contrast is nonlinear on scales $\ll R/8\Gamma$.  Indeed, our analysis in \S~\ref{sec:eom} suggests that nonlinear density contrast will be generated on scales $\lambda\lesssim  \xi_{\rm max}\sim \frac{1}{3}R/8\Gamma^2$, for which the expected number of vortical eddy turnovers is $\gg 3\Gamma$ (see \S~4.2 in \citealt{Sironi:07}).  On the other hand, since $\xi_{\rm max}$ is much larger than the plasma skin depth $c/\omega_{\rm p}$, the magnetic field that the turbulent dynamo generates will not be susceptible to the fast collisionless decay discussed by \citet{Chang:08} that affects the much smaller scale field that kinetic instabilities can generate in the shock transition layer \citep[e.g.,][and references therein]{Gruzinov:01,Frederiksen:04,Medvedev:05,Keshet:08,Spitkovsky:08}.

\subsection{Reversing Field from Bell's Streaming Instability}
\label{sec:bell}

In \S~\ref{sec:eom} we assumed that the magnetic field reverses direction on scales $\lambda$.  \citet{Bell:04,Bell:05} has shown that an instability driven by the nonthermal ion streaming can convert an initially uniform field into a reversing, helical magnetic field \citep[see, also,][]{Reville:06,Zirakashvili:08,Pelletier:08}.  For ultrarelativisic streaming and helical perturbations of the form $\xi_x+i\xi_y\propto \exp i(kR-\omega t)$ ($x$ and $y$ are two coordinates perpendicular to $\hat{\bf R}$), the instability has dispersion relation
\beq
\omega^2-k^2v_{\rm A}^2 - \frac{k J_{\rm ret} B_R}{\rho c}
\left(1-\frac{\omega}{kc}\right)=0 ,
\eeq
where $v_{\rm A}$ is the Alfv\'en velocity and $B_R\equiv {\bf B}\cdot{\hat{\bf R}}$.  The fastest growing mode has wavelength and growth rate
\beq
\label{eq:bell_fast}
\lambda_{\rm fast}=\frac{B_R}{en_{\rm ntp}}, \ \ \ \ \Upsilon_{\rm fast} = e n_{\rm ntp} \left(\frac{ n m_p}{\pi}\right)^{-1/2} ,
\eeq
respectively.\footnote{It is a peculiar characteristic of ultrarelativistic blastwaves that the gyroresonant wavelength is always longer than the particle's flight path in the shock upstream.}  For nonlinear density contrast production on scales $\lambda_{\rm fast}$ we need  $\xi(x_{\rm cool})> \lambda_{\rm fast}$ and $\xi(x_{\rm rel}) > \lambda_{\rm fast}$ as well as $\Upsilon_{\rm fast} > 8\Gamma^2 c/R$.  However, the wavelength of the fastest growing mode is very small if $B$ is the microgauss field typical of the interstellar medium.  All longer wavelengths $\lambda > \lambda_{\rm fast}$ also grow on rates
\beq
\label{eq:bell_rate}
\Upsilon(\lambda) \sim \Upsilon_{\rm fast} \left(\frac{\lambda}{\lambda_{\rm fast}}\right)^{-1/2} .
\eeq
This seems to suggest that a reversing field may be generated on a wide range of scales.

The instability has enough time to reach nonlinear growth only when the growth rate exceeds the inverse crossing time of the shock precursor
\beq
\label{eq:Upsilon_condition}
\Upsilon(\lambda) \gg  \left(\frac{R}{8\Gamma^2c}\right)^{-1}
\eeq
To evaluate the growth rate, we substitute equations (\ref{eq:density_of_delta}; coherent field) with $\Delta=R/8\Gamma^2$ and (\ref{eq:bell_fast}) in equation (\ref{eq:bell_rate}) to obtain
\beq
\Upsilon(\lambda) \sim \frac{2.4\ e^{1/2}c^{1/2} n^{1/4}  \epsilon_{B_0}^{1/4} \epsilon_{\rm nt}^{1/4}\Gamma^2}{\gamma_{p,{\rm max}}^{1/2} \Lambda^{1/2} m_p^{1/4} \lambda^{1/2}} .
\eeq
Where $\epsilon_{B_0}$ parametrizes the strength of the pre-existing coherent field, not to be confused with the same symbol expressing strength of the length scale-dependent tangled field in previous sections. The condition for nonlinear growth in equation (\ref{eq:Upsilon_condition}) is then fulfilled when
\bea
\lambda &\ll& \frac{R}{8\Gamma^2} \left(\frac{\gamma_{p,{\rm max}}}{4.6\times10^6}\right)^{-1}
\left(\frac{\epsilon_{B_0}}{10^{-9}}\right)^{1/2}
\left(\frac{\epsilon_{\rm nt}}{0.1}\right)
\left(\frac{\Gamma}{100}\right)^{4/3}\nonumber\\
& & \times
\left(\frac{n}{1\textrm{ cm}^{-3}}\right)^{1/6}
\left(\frac{E_{\rm tot}}{10^{53}\textrm{ erg}}\right)^{1/3} ,
\eea
which suggests that nonlinear growth is very broadly expected on scales $\lambda\lesssim \xi_{\rm max} \sim R/8\Gamma^2$ when $\gamma_{p,{\rm max}} \lesssim \gamma_{\rm rel,crit}$. When $\gamma_{p,{\rm max}}\gg \gamma_{\rm rel,crit}$, nonlinear growth is still expected on scales smaller than $R/8\Gamma^2$ but still much larger than the plasma skin depth.

\section{Discussion}
\label{sec:discussion}

The complexity of the structure of ultrarelativistic collisionless shocks stems from the multiscale nature of plasma self-organization.  This and the preceding analytical and numerical analysis have identified organization on various spatial scales between the proton plasma skin $c/\omega_{\rm p}\sim 2\times10^7\textrm{ cm}$ and the width of the blastwave in the upstream frame $R/8\Gamma^2\sim 10^{12}-10^{18}\textrm{ cm}$.  While the dynamical organization of the plasma on different scales is interdependent, our understanding of the scale interdependence in the precursors of ultrarelativistic collisionless shocks is still in its infancy. Motivated by the present and other preliminary investigations \citep[see, e.g.,][]{Milosavljevic:06,Katz:07,Keshet:08,Spitkovsky:08}, we propose several possible regions within the structure of an ultrarelativistic shock precursor in which the physical processes that drive plasma organization change character as a function of distance from the shock. We, however, caution that not all of these regions have to be present in all---or perhaps any---of the blastwave precursors.  Similarly, we are not able to determine the relative ordering of the regions, with respect to their distance from the shock transition, with any degree of certainty.  In a shock that produces a power-law spectrum of accelerated nonthermal particles (see \S \ref{sec:intro} for potential limitations of this key assumption), we propose that the shock precursor may contain the following concentric regions:

1. An outermost region extending to the maximum possible radial distance from the shock, measured in the reference frame of the shock upstream, of $\sim R/8\Gamma^2$, that is causally associated with the explosion. This region contains a low density of nonthermal protons in the high-energy tail of the nonthermal proton spectrum.  Because inverse Compton cooling losses prohibit the existence of nonthermal electrons in this region, a return current neutralizing the charge separation created by the high-energy accelerated nonthermal protons flows in the upstream plasma (\S~\ref{sec:charge_separation}). The effect of the return current is twofold: ({\it a}) starting from an initial, weak, coherent, quasi-uniform magnetic field, the return current nonresonantly excites circularly-polarized Alfv\'en waves, thereby converting the field component parallel to the direction of shock propagation into a wound, perpendicular, helical component (\S~\ref{sec:bell}); ({\it b}) given a perpendicular magnetic field, which could be pre-existing or generated by the return current, the Amp\`ere force associated with the return current accelerates the upstream plasma in the perpendicular direction (\S~\ref{sec:eom}) and produces a density contrast in the  plasma (\S~\ref{sec:dencon}).  More speculatively, it has been hypothesized that besides generating nonlinear density contrast, the return current interaction significantly amplifies the pre-existing magnetic field; this possibility awaits verification in numerical simulations.  Because the density of nonthermal particles in this region is low, the region is either stable to small-scale plasma instabilities, such as the transverse Weibel (or filamentation) instability \citep[e.g.,][]{Fried:59} and its relatives, or more likely, the small-scale instabilities are present but their growth times are longer than the crossing time of the region $\lesssim R/8\Gamma^2 c$.\footnote{Small scale instabilities may also occur of the upstream medium is enriched by an $e^\pm$-pair cascade \citep{RamirezRuiz:07}.} This region should exist unless the compression layer, which is the region 4 below, extends all the way to the edge of the maximum radial distance from the shock ($\sim R/8\Gamma^2$), which renders the shock transition layer as wide as the downstream.

2. A region still devoid of nonthermal electrons and endowed with the return current, but where the density of nonthermal protons is high enough to drive the small-scale plasma instabilities, with wave vectors $k\sim \omega_{\rm p}/c$.  At present, the dominant unstable kinetic mode in the regime in which a very low density beam of particles with an ultrarelativistic forward momentum and an ultrarelativistic transverse momentum spread streams through a cold background plasma has not been identified.  It is possible that the small-scale instabilities can generate an electromagnetic field efficiently and that the field enhances the perpendicular velocity diffusion of nonthermal particles, which is a necessary element of the DSA cycle.  The magnetic field in this region is a combination of a small-scale field generated by the instabilities and a larger scale field generated from the pre-existing field by the return current.  If the latter is strong enough, depending on its resulting orientation relative to the shock normal, it may suppress some of the small-scale instabilities, or it may drive small-scale turbulence in the flow, especially if the resulting field is perpendicular to the shock normal \citep{Hededal:05}. 

3. A region close enough to the shock transition that the inverse Compton cooling time of the nonthermal electrons is longer than the duration of the DSA cycle.  Since this region is populated by nonthermal electrons and protons, the nonthermal particle precursor is neutral overall, and no return current is present.  However, small scale instabilities are still present and are more intense than in the preceding regions.  The small scale instabilities generate a strong magnetic field in the upstream plasma, which, when advected to the shock transition, facilitates upstream particle scattering and thermalization.  The small scale field may also play a critical role in the injection and acceleration of particles in the DSA mechanism \citep[e.g.,][]{Lemoine:06b}. The relative ordering of regions 2 and 3 may vary. Depending on the parameters and initial conditions, it may happen that the region containing nonthermal electrons extends farther from the shock than that in which small scale plasma instabilities excited by nonthermal particles play an important role in setting the shock structure.

4. The shock transition region, where the upstream plasma is compressed. At the leading edge of this transition layer, the compression may result from an efficient momentum transfer from the nonthermal particles \citep[e.g.,][]{Blandford:80,Drury:81}, which could be coupled to the upstream plasma by the small scale instabilities. In this region, the upstream plasma starts to accelerate (in the rest frame of the far upstream) and decelerate (in the rest frame of the shock).  The final stage of compression, where the upstream becomes Newtonian in the rest frame of the shock, takes place  when the upstream interpenetrates the downstream isotropized plasma, and so the hydrodynamic jump conditions are realized. If the nonthermal particles play a role in initiating the compression, than the width of this layer is much larger than the plasma skin depth. If, however, efficient compression commences only where the upstream interpenetrates the downstream thermalized plasma, the width of the compression layer may be as small as of the order of the skin depth. While in the preceding regions we expect a decrease in the mean magnetic field correlation length $\langle\lambda\rangle\equiv \langle\lambda \bar B^2(\lambda)\rangle/\langle \bar B^2(\lambda)\rangle$ as the plasma approaches the shock, in the shock transition region we expect an opposite trend: the small scale instabilities have reached saturation and the small field they have generated decays via a kinetic damping mechanism as the shocked plasma travels downstream \citep{Gruzinov:01,Medvedev:05,Chang:08,Keshet:08,Spitkovsky:08}.  As the small scale field decays, the residual field, which partakes in the synchrotron emission of the GRB afterglows \citep[see, e.g.,][]{Panaitescu:02,Yost:03} and must be much stronger than the shock compressed pre-existing mangetic field of the normal interstellar medium \citep[see, e.g.,][]{Gruzinov:01,Piran:05b}, is the larger scale field generated in regions 1-3 above. The shock transition region is also the site of the onset of the establishment of electron-ion equipartition \citep{Spitkovsky:07}.

Finally we note that if the highest energy nonthermal particles traverse regions 1-3 during a DSA cycle, that is, these particles are deflected by the magnetic field of the non-accelerated upstream, then an upper limit on $\gamma_{p,{\rm max}}$ can be set \citep{Milosavljevic:06}. This limit is derived from the constraint of causality on the self organization scale of the upstream magnetic field ($\sim R/8\Gamma^2$) and the maximal magnetic field energy density in the non accelerated upstream ($\sim nm_pc^2$). This upper limit is, roughly, $\gamma_{p,{\rm max}} \lesssim 10^{10}\ n^{1/2}\ (R/10^{18} \textrm{ cm})$ implying the GRB external shocks are not promising sources of ultra-high energy cosmic rays (UHECR) with Lorentz factors, under the assumption that they are protons, of $\gamma_p\gtrsim 10^{11}$.

\section{Conclusions}
\label{sec:conclusions}

We have examined the influence of a pre-existing magnetic field on the structure of ultrarelativistic shock waves, such as in the external GRB shocks that may be behind the GRB afterglow phenomenon.  From the hypothesis that the shock wave accelerates charged particles to high Lorentz factors, we were able to derive the following conclusions:

In GRB external shocks that accelerate ions to Lorentz factors $\gamma_i\gg 10^3$, inverse Compton cooling in the shock upstream hinders nonthermal electron acceleration to high energies.  Because of this, nonthermal electrons do not penetrate as far into the shock upstream as the highest-energy nonthermal ions do.  Thus, a region exists in the shock upstream far from the shock transition at $R\sim ct[1-(8\Gamma^2)^{-1}]$, but still behind the photon shell at $R=ct$, where return current flows to cancel the charge separation between the nonthermal electrons and ions.

As the return current flows across the weak pre-existing magnetic field of the shock upstream, the Amp\`ere force accelerates the upstream plasma perpendicular to the direction of shock propagation.  If the upstream field reverses on scales $\lambda\ll R/8\Gamma^2$, the transverse acceleration produces a nonlinear density contrast on the same scales.  If the pre-existing field is quasi-uniform (non-reversing), Bell's streaming instability will efficiently convert the quasi-uniform field into a reversing field during the light-crossing time of the shock precursor.

As the shock transition sweeps the resulting density clumps, the bulk kinetic energy of the blastwave is converted into vortical energy.  The vortical eddies  in the shock downstream  turn over many times and  amplify the magnetic field via a dynamo mechanism.  For clumps of peak density contrast of about one, the vortical energy \citep{Sironi:07} and with it the magnetic energy density can be amplified to $\epsilon_B \sim (3\Gamma)^{-1}$, where in GRB external shocks, $\Gamma \lesssim  \textrm{ few }\times100$. This is consistent with the observed GRB light curves and spectra. The generated magnetic field will be coherent on scales much larger than the plasma skin depth and will, as such, not be susceptible to rapid collisionless decay.
Taken together, these results indicate that the microgauss magnetic field ($\epsilon_B\sim10^{-9}$) typical of the interstellar medium is sufficient to drive Bell's mechanism in ultrarelativistic GRB external shocks and deliver nonlinear density inhomogeneities to the shock transition.  \emph{Therefore, magnetic field amplification through shock vorticity generation is expected even when the circumburst medium is not initially clumpy, and we expect it to operate in long and short GRBs.}

We speculatively propose a structure for the shock precursor consisting of four concentric zones characterized by different forms of plasma self-organization.  First, the outermost region of the precursor is the region populated by streaming, nonthermal protons but devoid of nonthermal electrons.  There, the plasma develops non-linear density inhomogeneities.  Second, a region that still contains a return current but where the nonthermal particle precursor streaming through the shock upstream is susceptible to small-scale kinetic plasma instabilities.  Third, a region containing the relatively low-energy nonthermal electrons that can be accelerated in the shock in spite of the inverse Compton cooling losses; in this region, the return current is absent.  Finally, the fourth region is the shock transition itself, where kinetic instabilities and plasma waves facilitate plasma isotropization and establish the hydrodynamic jump. The relative ordering of the second and the third region may vary.

\acknowledgements

We thank the anonymous referee for emphasizing uncertainties associated with the efficiency of the diffusive shock acceleration process in ultrarelativistic collisionless shocks, and acknowledge numerous invaluable discussions with Uri Keshet and Anatoly Spitkovsky.  This research was supported by a Senior Research Fellowship from the Sherman Fairchild Foundation and NASA grant NNH05ZDA001N (to E. N.).

\end{document}